\documentclass[11pt,preprint]{aastex}

\usepackage{amsmath,amstext}
\usepackage[T1]{fontenc}
\usepackage{graphicx}
\usepackage{amsmath,amssymb}
\usepackage{color}
\usepackage{bm}
\usepackage[breaklinks,colorlinks,citecolor=blue,linkcolor=blue]{hyperref} 
\usepackage[all]{hypcap} 
\usepackage[normalem]{ulem}
\usepackage{soul}
\bibpunct{(}{)}{;}{a}{}{,}

\def\msun{M_{\odot}}
\def\kms{\rm km\,s^{-1}}
\def\cmc{\rm cm^{-3}}

\def\ns{n_{\rm s}}

\def\ts{T_{\rm s}}

\def\tdyn{t_{\rm dyn}}
\def\tff{t_{\rm ff}}
\def\tcool{t_{\rm cool}}
\def\tvir{T_{\rm vir}}
\def\rvir{R_{\rm vir}}
\def\tc{T_{\rm c}}

\def\fsn{f_{_{\rm SN}}}
\def\esn{E_{_{\rm SN}}}
\def\wsn{\omega_{_{\rm SN}}}
\def\qimf{q_{_{\rm IMF}}}

\def\estar{\epsilon_{\star}}
\def\zc{z_{\rm c}}

\def\rhos{\rho_{\star}}
\def\rhog{\rho_{\rm g}}
\def\vesc{v_{\rm esc}}
\def\dcj{D_{\rm CJ}}
\def\tstar{t_{\star}}
\def\tcool{t_{\rm cool}}
\def\tdyn{t_{\rm dyn}}
\def\mcl{M_{\rm cl}}
\def\rcl{R_{\rm cl}}

\def\tff{t_{\rm ff}}
\def\ncl{n_{\rm cl}}
\def\cs{c_{\rm s}}
\def\rf{r_{\rm f}}
\def\zf{z_{\rm f}}

\def\uf{u_{\rm f}}
\def\vf{v_{\rm f}}
\def\zp{z^{\prime}_{\rm f}}
\def\zpp{z^{\prime\prime}_{\rm f}}

\title{Self-sustaining star formation fronts in filaments during cosmic dawn}
\author{Xiawei Wang and Abraham Loeb}
\affil{Department of Astronomy, Harvard University, 60 Garden Street, Cambridge, MA 02138, USA}%

\begin{abstract}
We propose a new model for the ignition of star formation in low-mass halos by a self-sustaining shock front in cosmic filaments at high redshifts.
The gaseous fuel for star formation resides in low mass halos which can not cool on their own due to their primordial composition and low virial temperatures.
We show that star formation can be triggered in these filaments by a passing shock wave.
The shells swept-up by the shock cool and fragment into cold clumps that form massive stars via thermal instability on a timescale shorter than the front's dynamical timescale.
The shock, in turn, is self-sustained by energy injection from supernova explosions.
The star formation front is analogous to a detonation wave, which drives exothermic reactions powering the shock.
We find that sustained star formation would typically propel the front to a speed of $\sim 300-700\,\kms$ during the epoch of reionization. 
Future observations by the $\textit{James Webb Space Telescope}$ could reveal the illuminated regions of cosmic filaments, and constrain the initial mass function of stars in them.
\end{abstract}
\begin{document}
\section{Introduction}
The gas reservoir of low-mass halos at high redshifts exhibits inefficient star formation due to the lack of metals, which are essential for the transition from intermediate temperature atomic gas to cold molecular gas \citep{krumholz2012, loeb2013}.
Nevertheless, a significant population of star-forming galaxies beyond $z\gtrsim10$ is required to explain the Thomson optical depth of the cosmic microwave background \citep{finkelstein2015, robertson2015}.
A likely compensating factor for the shortage of ionizing photons is a population of faint low-mass halos \citep{bouwens2012, anderson2017}, observationally suggested by the steep faint end slope of the UV luminosity function \citep{finkelstein2015, anderson2017}.
The process by which efficient star formation is initiated in low-mass halos at high redshifts is still unknown, given the inefficient star formation rate (SFR) observed in low-mass halos at low redshifts \citep{behroozi2013}.
Therefore, it is important to probe the SFR in low-mass halos during the epoch of reionization through future observations with the $\textit{James Webb Space Telescope}$ ($JWST$).

Galactic outflows play an important role in the formation and evolution of low-mass galaxies (e.g. \citealt{dekel1986, peeples2011}), as well as in regulating star formation \citep{silk1997, hopkins2011} and the enrichment of circumgalactic and intergalactic medium \citep{furlanetto2003}.
Cold molecular clouds are identified in observations of such outflows \citep{rupke2005, sturm2011}.
Numerical simulations have shown that outflowing shells tend to fragment through a thermal instability \citep{thompson2016, ferrara2016, scannapieco2017, schneider2018}, which may lead to subsequent star formation within the outflows \citep{silk2013, zubovas2014, maiolino2017, wang2018}.
However, previous studies of galactic outflows were limited to the scale of the host galaxy and the surrounding circumgalactic and intergalactic medium. 
How these outflows may affect their neighboring halos remained unclear.

In this work, we propose a new model for the ignition of star formation in low-mass halos that otherwise do not form stars.
Such halos are often distributed in filaments.
A passing shock could trigger star formation and generate a self-sustaining starburst front.
We make an analogy between this process and the propagation of a detonation wave, in that the gas reservoir of low-mass halos is analogous to gunpowder, and the burning front triggers new star formation while being dynamically maintained by the energy release from supernovae (SNe).
The paper is organized as follows.
In \S\ref{sec:section2}, we describe our model in analogy to detonation wave theory.
In \S\ref{sec:section3}, we calculate the propagation of the star formation front and present numerical results.
Finally, \S\ref{sec:section4} summarizes our main results and observational implications.

\section{Star formation front}
\label{sec:section2}
\begin{figure}[h!]
\includegraphics[angle=0,width=\columnwidth]{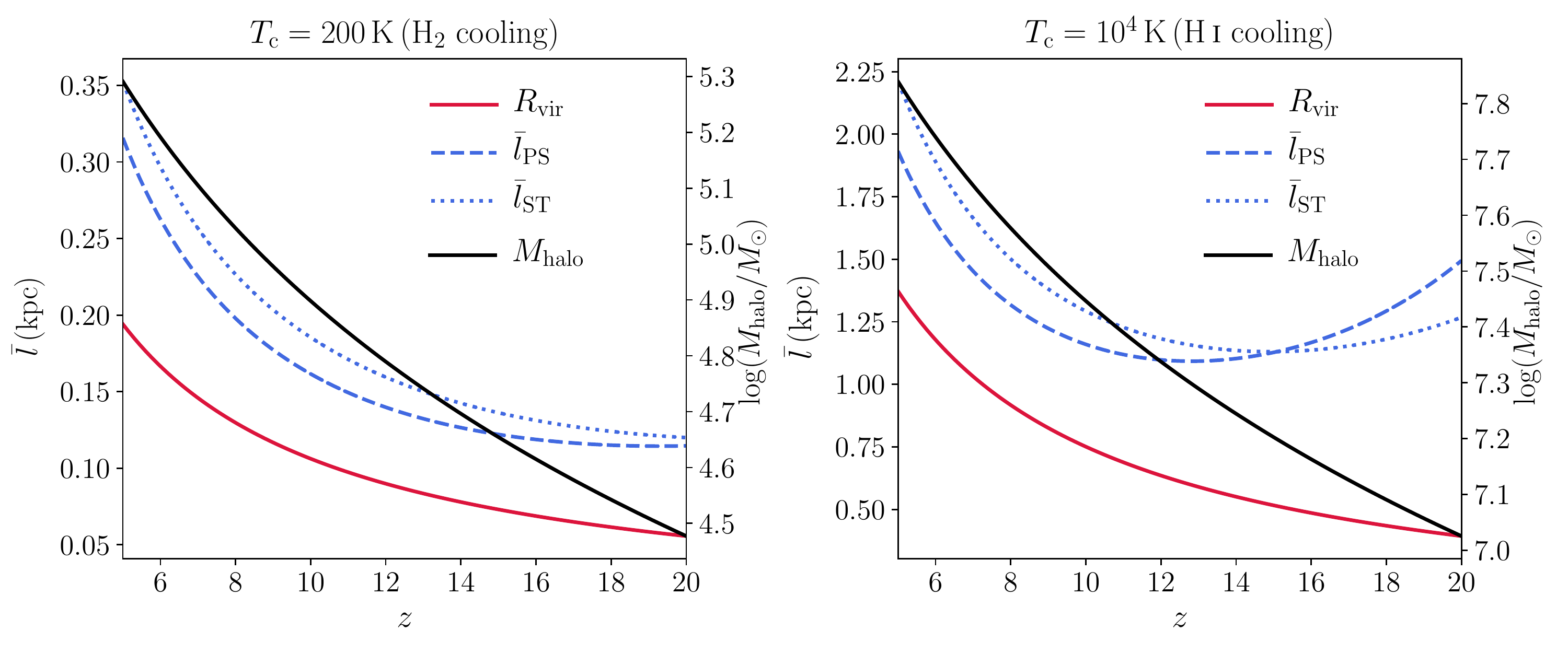}
\caption{
Comparison of the average separation of halos with $\tvir=0.5-1.0\,\tc$, and their $\rvir$, in cosmic filaments.
Two panels show the cases of H$_2$ and H$\textsc{i}$ cooling thresholds, respectively.
The red solid lines represent $\rvir$, while the dashed and dotted blue lines correspond to the average separation estimated from the Press-Schechter \citep{press1974} and Sheth-Tormen \citep{sheth1999} halo mass functions, respectively ($\sim10$ times more compact in filaments).
The black lines provide the halo mass, $M_{\rm halo}$, whose $\tvir$ is just below $\tc$, with the scale labeled on the right-hand vertical axes.
This implies that halos just below $\tc$ are tightly packed in filaments at high redshifts.
}
\label{fig:fig1}
\end{figure}

The gas reservoir of low-mass halos can not initiate star formation on its own and remains quiescent if the virial temperature of the halo, $\tvir$, is below the cooling threshold temperature, $\tc$.
For primordial gas composition, the cooling threshold can be at minimum $\tc=200$ K for molecular hydrogen, H$_2$ (below which molecular transitions are not excited) or $\tc=10^4$ K for atomic hydrogen, H$\,\textsc{i}$, if H$_2$ is dissociated by a UV background (see review in Chapter 6 of \citealt{loeb2013}).
Prior to star formation and feedback, sufficient baryons have been assembled into these halos as their virial masses exceed the cosmological Jeans (filtering) mass at $z\gtrsim20$ \citep{haiman1996}.
The halos experience a gas-poor phase when feedback partially removes the gas, but recover a gas-rich phase when the gas accretes back from the IGM.
The recycling of baryons results in an average baryon fraction which is $\sim 50\%$ of the cosmic average in halos of masses $\sim 10^7\,\msun$, with a lower fraction in lower-mass halos \citep{chen2014, wise2014}.
However, during gas-rich phases, halos maintain a baryon fraction that is approximately the cosmic average \citep{chen2014}.
The average separation of halos with $\tvir$ in the range of $\sim 0.5-1.0$ $\tc$ is $\bar{l}\approx (4\pi n/3)^{-1/3}/(1+z)$, where $n$ is the comoving number density of dark matter halos as derived from the halo mass function \citep{press1974, sheth1999} .
In cosmic filaments, halos are found to be closer together \citep{bond1996, mo2010}, with $\bar{l}$ smaller by up to a factor of $\sim10$.
Figure \ref{fig:fig1} shows that $\bar{l}$ is a few times $\rvir$, and should shrink to $\sim\rvir$ inside filaments. 
Thus, we assume that halos are contiguous with their neighbors, tightly packed along the filament.

\begin{figure}[h!]
\centering
\includegraphics[angle=0,width=0.7\columnwidth]{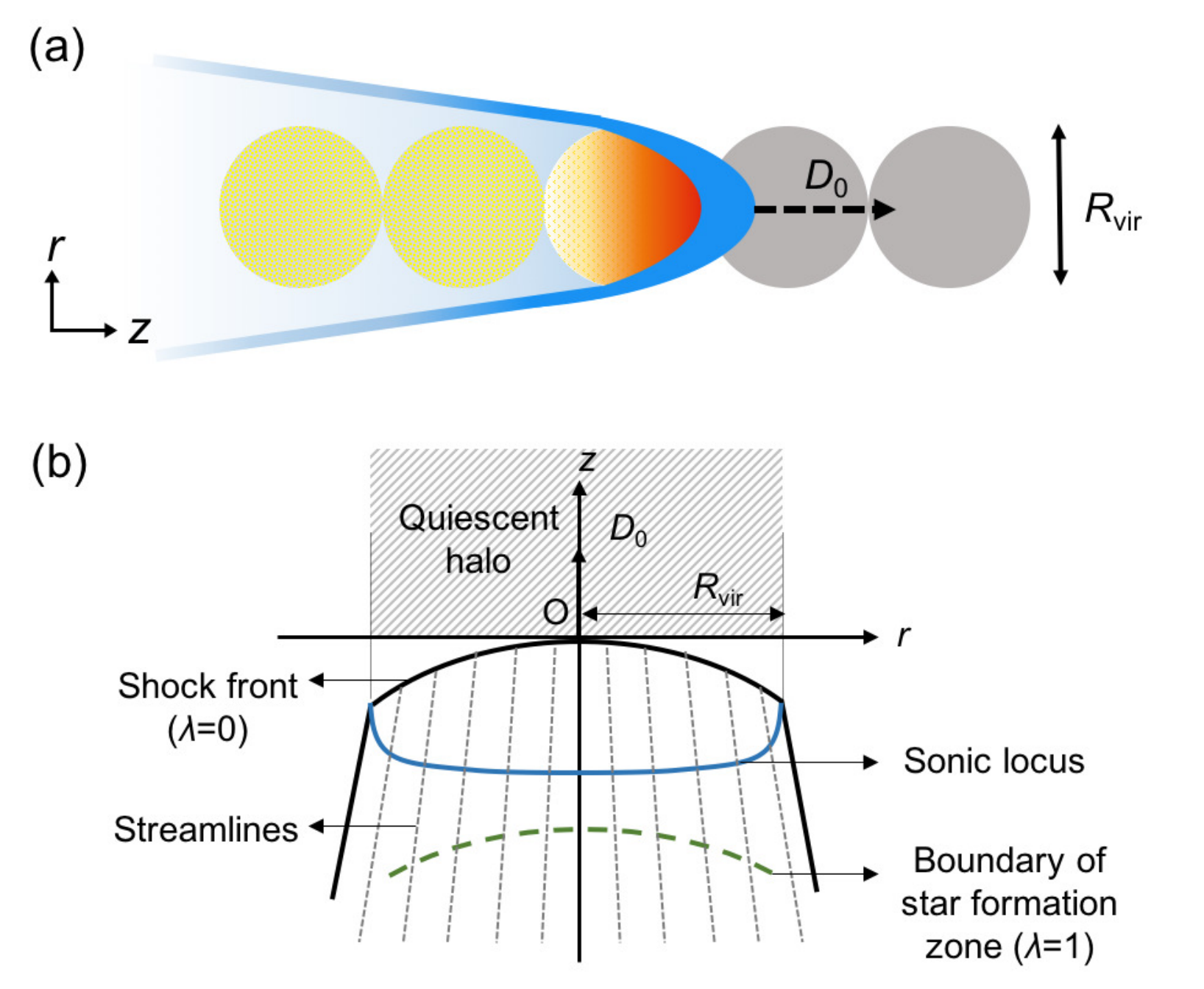}
\caption{
Configuration of a star formation front.
Panel (a) sketches the burning front sweeping through halos packed in a filament that can not form stars before the passage of the shock, analogous to a self-propagating detonation wave in gunpowder.
The width of the filament is $\sim\rvir$.
Panel (b) shows the schematic diagram of a two-dimensional cylindrical detonation wave.
The blue solid line represents the sonic locus, where the Chapman-Jouguet (CJ) condition is satisfied.
The green dashed line shows the boundary of the star formation zone, behind which SF had completed.
The grey dotted lines correspond to the streamlines of post-shock flows, which are assumed to be straight but diverging (see Appendix for details).
Ahead of the detonation shock front lies the unburnt fuel of low-mass halos, in which stars can not form until the shock's passage.
}
\label{fig:fig2}
\end{figure}

A galactic outflow driven by active galactic nuclei or SN would propagate supersonically and sweeps up the ambient medium with a speed of hundreds of $\kms$, as based on observations and theoretical calculations (e.g. \citealt{king2015, wang2015}).
The shells swept-up by the outflow tend to cool rapidly and fragment into cold clumps that subsequently form stars \citep{zubovas2014, scannapieco2017, wang2018}.
The outflow shock is rejuvenated as it gains energy from new SN explosions which sustain its propagation.
Hence, star formation can be ignited by the front as it passes through the filament.
The configuration of such a burning front of star formation is shown in Fig. \ref{fig:fig2}.
The propagation of this self-sustaining shock is analogous to a detonation wave, which involves an igniting shock self-sustained by an exothermal chemical process \citep{fickett1979}. 
%
\subsection{Detonation model}
The reactive Euler equations of high-speed flows coupled to energy release can be used to describe the propagation of star formation fronts, in analogy to detonation waves.
These equations are,
\begin{subequations}
\label{eq:subeqns}
\begin{gather}
\frac{D\rho}{Dt}+\rho\nabla\cdot\bm{v}=0\;\,
\label{eq:subeq1}\\
\rho\frac{D\bm{v}}{Dt}=-\nabla p\;\,
\label{eq:subeq2}\\
\frac{De}{Dt}-\frac{p}{\rho^2}\frac{D\rho}{Dt}=0\;\,
\label{eq:subeq3}\\
\frac{D\lambda}{Dt}=W\;\,
\label{eq:subeq4}
\end{gather}
\end{subequations}
where $D/Dt=\partial/\partial t+\bm{v}\cdot\nabla$ is the full time derivatives of the flow.
In a steady-state, $\partial/\partial t=0$ in the rest frame of the detonation wave.
Throughout our discussion, $\bm{v}$, $\rho$ and $p$ are the velocity, density and pressure of the flow, respectively; $e$ is the internal energy per unit mass; $\lambda=\rhos/(\rhos+\rhog)$ is the stellar mass fraction; and $\rhos$ and $\rhog$ are the stellar and gas density, respectively.
The location where $\lambda=0$ corresponds to the detonation shock front, whereas $\lambda=1$ corresponds to the completion of star formation.
We approximate the fuel distribution as uniform, i.e. smooth over the scale of individual halos when describing the global propagation of the front.
This approach is similar to the description of detonation waves in gunpowder, which involves smoothing over the scale of individual grains in the fuel.
We adopt the polytropic equation of state:
\begin{equation}
e=\frac{p}{(\gamma-1)\rho}-Q\lambda\;,
\end{equation}
where $\gamma=5/3$ is the polytropic index.
Here $Q$ is the energy release from SN explosions per unit mass of gas:
\begin{equation}
Q=\frac{\qimf\fsn\esn}{\wsn}\;,
\end{equation}
where $\esn$ is the energy released by each SN, $\wsn$ is the total amount of stellar mass that must be formed in order to produce one SN.
For a very massive initial mass function (IMF), $\esn=10^{52}$ ergs and $\wsn=462\,\msun$ \citep{furlanetto2003}.
The coefficient $\fsn\sim0.25$ is the fraction of the energy produced by SN to power the wind while the rest is lost mainly to radiative cooling (e.g, \citealt{mori2002}).

The parameter $\qimf$ quantifies a deviation of the IMF from Pop III stars.
In Eq. (\ref{eq:subeq4}), $W\equiv d\lambda/dt$ denotes the SFR, derived from the Kennicutt-Schmidt (KS) law \citep{kennicutt1998} and converted to the volume density of SFR \citep{schaye2008}:
\begin{equation}
\dot{\rhos}=A^{\prime}(1\,\msun\,\textrm {pc}^{-2})^{-n^{\prime}}\left[\frac{\gamma}{\it G}(1-\lambda) p\right]^{(n^{\prime}-1)/2}\rhog\;,
\end{equation}
where $G$ is Newton's constant; $A^{\prime}=(2.5\pm0.7)\times10^{-4}$ and $n^{\prime}=(1.4\pm0.15)$ are the normalization constant and power-law index in KS law for surface density.
Thus,
\begin{equation}
W=\estar\frac{\dot{\rhos}}{\rhos+\rhog}=\estar A p^n(1-\lambda)^m\;,
\end{equation}
where $A$ is a normalization constant.
We adopt $n^{\prime}=1.5$ and derive the power-law indices $n=0.25$ and $m=1.25$.
Here, $\estar$ is a correction factor for the formation rate of Pop III stars which could be different from KS law due to their low metallicity \citep{trenti2009}. 
The fiducial values of free parameters are: $\fsn=0.25$, $\qimf=1.0$, $\estar=1.0$, $\esn=10^{52}$ erg, $\wsn=462\,\msun$, $m=1.25$ and $n=0.25$.

\begin{figure}[h!]
\includegraphics[angle=0,width=\columnwidth]{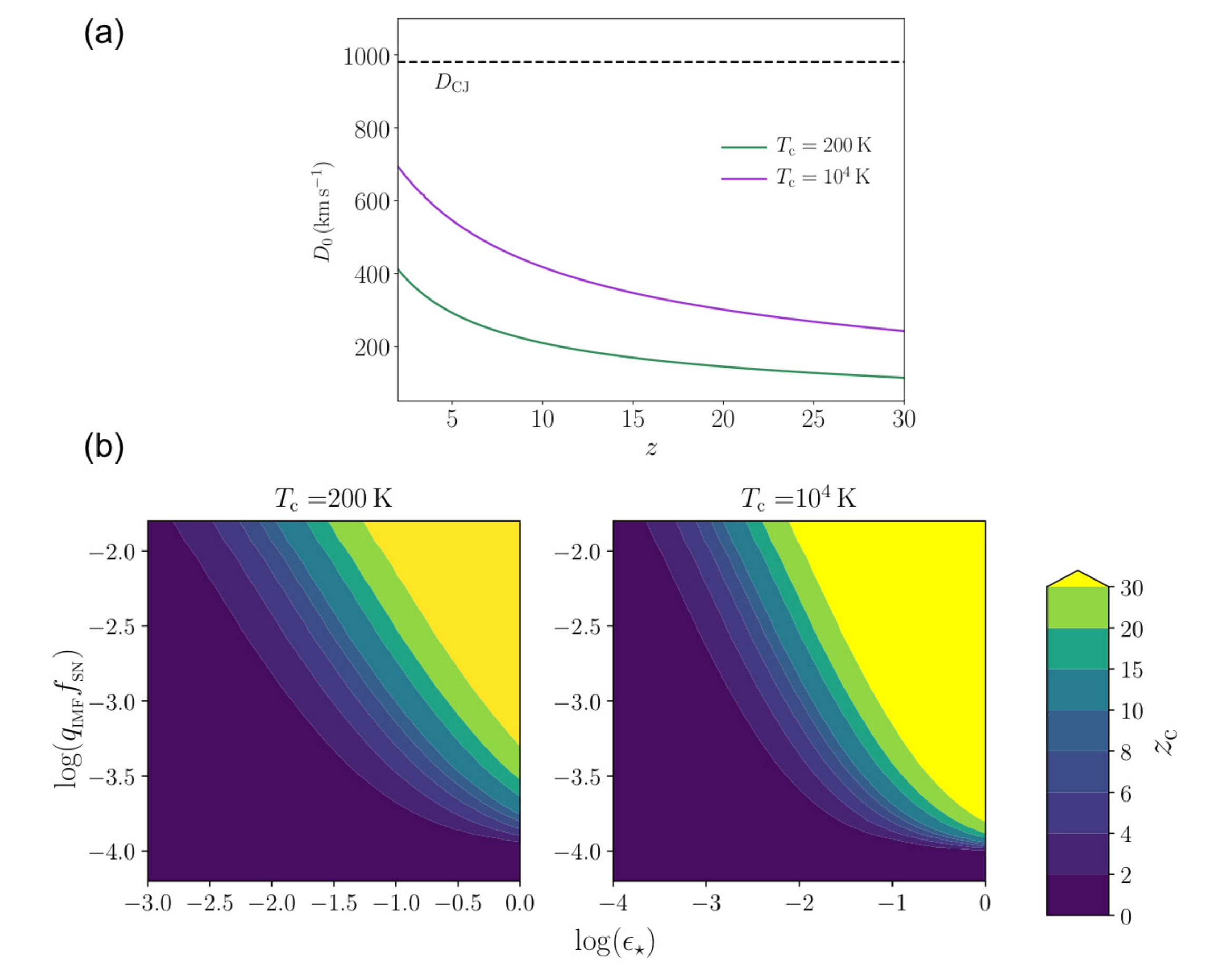}
\caption{
Panel (a) shows the average detonation speed, $D_0$, as a function of $z$, for $\tc=200$ K  (green) and $10^4$ K (purple).
The dashed line represents the ideal CJ speed, $\dcj\approx10^3\,\kms$.
Model parameters are taken to be their fiducial values.
Panel (b) represents the critical redshift of detonation failure, $\zc$, as a function of model parameters $\estar$, $\qimf$, $\fsn$.
We show contours of the $\zc$, beyond which the detonation fails, for $\tc=200$ K (left) and $\tc=10^4$ K (right).
The horizontal and vertical axes span different values of $\estar$ and the product of $\qimf$ and $\fsn$, respectively.
We fix other parameters to their fiducial values.
The color bar indicates $\zc$.
}
\label{fig:fig3}
\end{figure}

To solve Eqs. (\ref{eq:subeq1})-(\ref{eq:subeq4}), we follow the semi-analytical approach from \citet{watt2012} (see Appendix for details).
In particular, we find that the average steady-state detonation speed, $D_0$, decreases as $\rvir$ decreases, or equivalently, as $z$ increases, as shown in Fig. \ref{fig:fig3}.
The star formation front travels with a speed of $D_0\sim 200-400\,\kms$ at the beginning of reionization ($z\sim 30$), and $D_0\sim 300-600\,\kms$ at the end of reionization ($z\sim6$).
We show that $D_0$ is a fraction, $\sim0.2-0.7$ of the idealized one dimensional detonation speed, $\dcj=[2(\gamma^2-1)Q]^{1/2}\approx1000\,\kms$, for the free parameters set at their fiducial values. 
The significant deviation from a one-dimensional solution indicates that lateral expansion and energy losses along radial direction are non-negligible, particularly for high-redshift halos with a smaller $\rvir$.
%
\subsection{Star formation}
Next, we follow the prescription of our previously derived model for star formation within shells \citep{wang2018}.
The swept-up gas cools and fragments into cold clumps embedded in a hot tenuous gas via a thermal instability, which occurs if the heating rate at a constant pressure rises faster than the cooling rate as a function of temperature, consistent with observations \citep{maiolino2017} and numerical simulations \citep{ferrara2016, schneider2018}.
The cooler gas continues to condense at a constant pressure, leading to the formation of a two-phase medium \citep{field1965, silk2013, zubovas2014, inoue2015}.
The cooling timescale of the swept-up gas can be estimated as $\tcool\approx3.3\times10^3\, n_1^{-1} T_4\Lambda_{-23}^{-1}(T,Z)$ yrs, where $n_1=(\ns/1\,\cmc)$ is the number density of post-shock gas, $T_4=(\ts/10^4\,\rm K)$ is the post shock gas temperature, $\Lambda_{-23}=(\Lambda/10^{-23}\,\rm erg\,cm^3\,s^{-1})$ is the cooling function, and $Z$ is the metallicity (e.g. \citealt{maio2007, arata2018}).
For halos of mass $\sim10^8\,\msun$ and size $\sim0.5$ kpc, the characteristic shocked gas density at redshift $z\sim10$ is $\ns\sim10\,\cmc$.
For $Z\lesssim10^{-2}Z_{\odot}$, where $Z_{\odot}$ denotes solar metallicity, $\Lambda_{-23}$ is in the range $10^{-3}-0.1$, approximately scaling as $\sim(Z/Z_{\odot})$ \citep{sutherland1993, maio2007, inoue2015}.
Thus, $\tcool$ is much shorter than the dynamical timescale of the flow, $\tdyn\sim\rvir/D_0\sim10^7$ yrs.
The characteristic mass and size of the clouds induced by the thermal instability can be estimated as $\mcl\sim110\,T_{\rm h,6}^{9/2}n_{\rm h,0}^{-2}\,\msun$ and $\rcl\sim0.22\,T_{\rm cl,1}^{1/3}T_{\rm h,6}^{7/6}n_{\rm h,0}^{-1}$ pc \citep{field1965, wang2018}, where $T_{\rm h,6}=(T_{\rm h}/10^6\,\rm K)$ and $n_{\rm h,0}=(n_{\rm h}/1\,\cmc)$ are the temperature and number density of the hot medium embedding the clouds, and $T_{\rm cl,1}=(T_{\rm cl}/10\,\rm K)$ is the temperature of the clouds.
The gas clouds induced by thermal instability have a particle number density of $n_{\rm cl}\sim 10^4\,\cmc$, and will therefore collapse to form stars on a free-fall timescale $\tff\sim(G\rho)^{-1/2}\sim10^6\,n_{\rm cl,4}^{-1/2}$ yrs $\ll\tdyn$, where $n_{\rm cl,4}=(\ncl/10^4\,\cmc)$.

%
\section{Numerical results}
\label{sec:section3}
\begin{figure}[h!]
\centering
\includegraphics[angle=270,width=0.8\columnwidth]{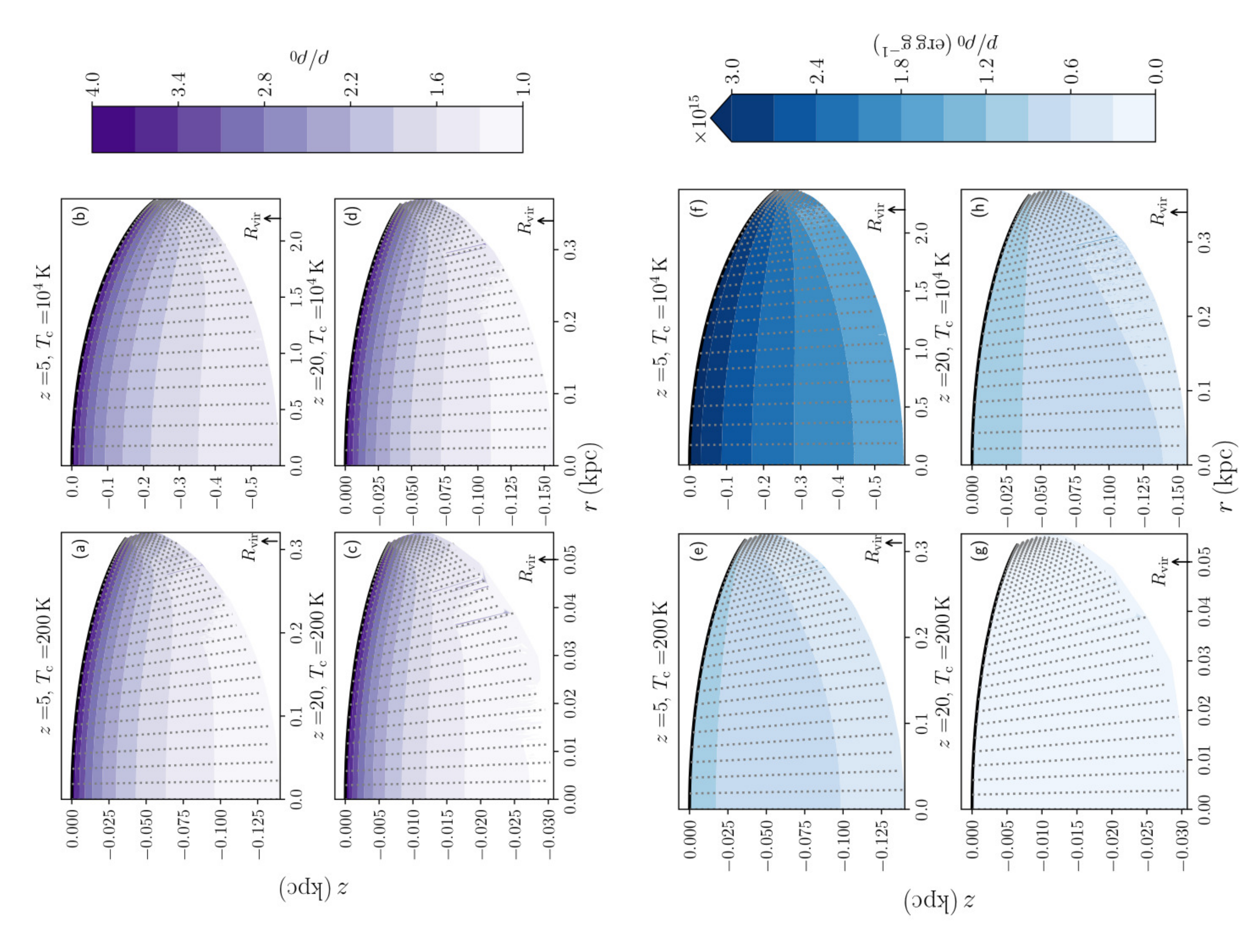}
\caption{
Flow density and pressure behind the detonation shock.
We show the ratio of flow density (panel (a)-(d)) and pressure (panel (e)-(h)) behind the shock to the ambient medium density $\rho_0$, for $z=5,\,20$ and $\tc=200$ K and $10^4$ K.
In each panel, the solid black and dashed grey curves represent the shock front and the flow streamlines, respectively.
The end of the streamlines marks the sonic locus.
The values of $\rvir$ are shown at the bottom right corner.
}
\label{fig:fig4}
\end{figure}

\begin{figure}[h!]
\centering
\includegraphics[angle=270,width=0.8\columnwidth]{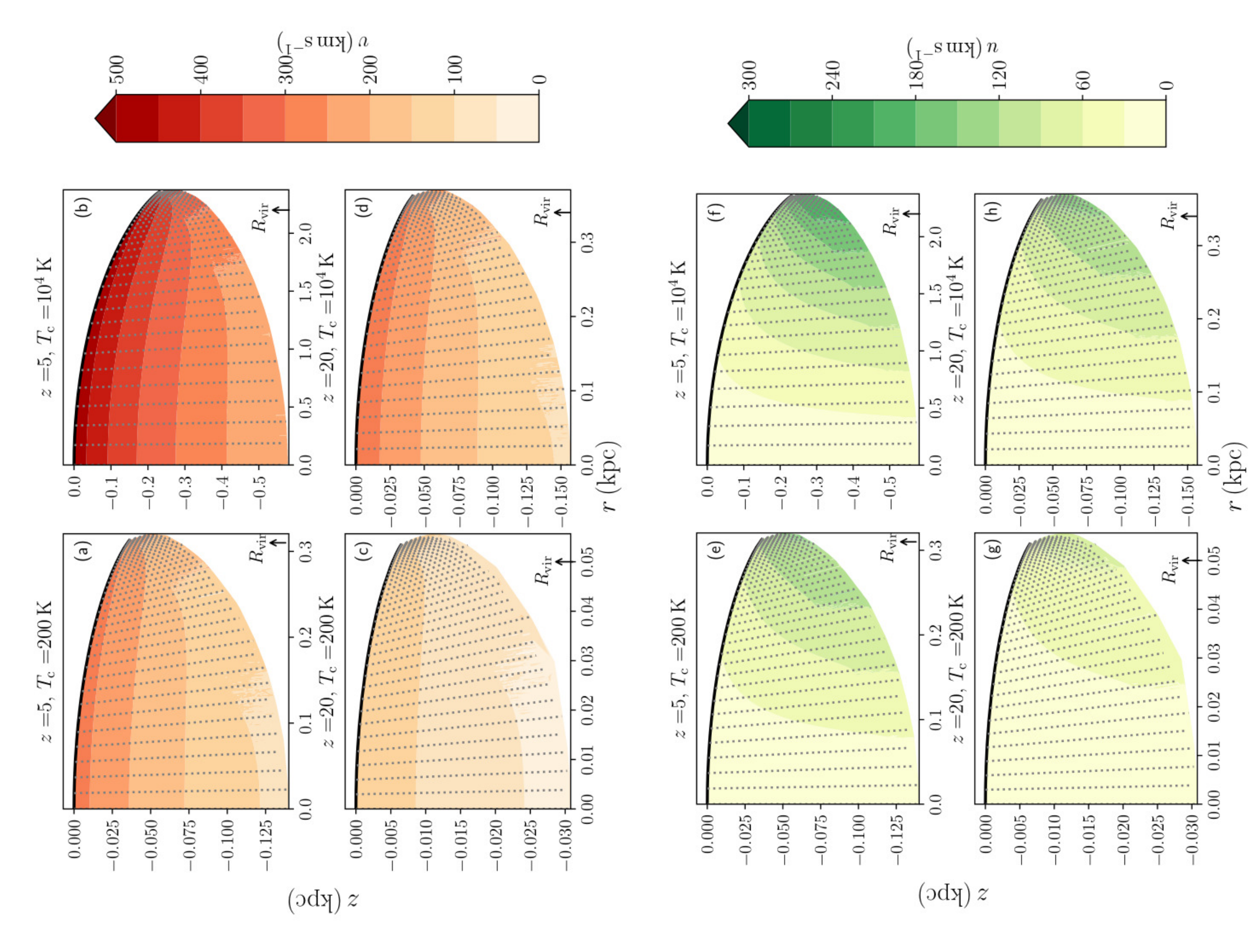}
\caption{
Axial flow speed (panel (a)-(d)) and radial flow speed (panel (e)-(h)), behind the detonation shock.
The arrangement of the plot is the same as in Fig. \ref{fig:fig4}.
The detonation wave experiences energy loss due to lateral expansion, and thus $D_0$ departs from $\dcj$.
}
\label{fig:fig5}
\end{figure}

We note that $D_0$ must exceed the maximum of the local sound speed, $\cs$, and the escape speed of the halo, $\vesc$, in order to remain supersonic and capable of entering neighboring halos.
Figure \ref{fig:fig3} shows the critical redshift, $\zc$, beyond which the detonation mode of the star formation front fails to satisfy this requirement.
Overall, we find that the star formation front is self-sustainable for a broad range of $\qimf\fsn$ and $\estar$.
We numerically solve Eq. (\ref{eq:subeqns}) for the density, pressure, axial and radial velocities of the flow behind the star formation shock front, as shown in Figures \ref{fig:fig4}-\ref{fig:fig5}, for model parameters taken at their fiducial values.
We find that two-dimensional effects are more significant in halos with smaller radii, which suffer from energy losses due to lateral expansion.
In these halos, the star formation front propagates with a moderate speed of $\sim300\,\kms$, while in halos with a larger $\rvir$, $D_0$ reaches $\gtrsim700\,\kms$.
The star formation front is curved due to lateral expansion and the streamlines in the flow diverge.
In the rest frame of the star formation front, the sonic locus, shown as the lower boundary in Figures \ref{fig:fig4}-\ref{fig:fig5}, is the place where the flow speed is equal to the local sound speed.
Star formation and energy release are incomplete in the subsonic zone between the shock front and sonic locus, behind which the flow is supersonic in the detonation front rest frame.
Therefore, only the energy injection from this region, also known as the detonation driving zone \citep{watt2012}, is available to drive the propagation of the star formation front.
$D_0$ is less than the ideal one-dimensional value, $\dcj$, and depends on the shock curvature and $\rvir$, consistently with the results shown in Fig. \ref{fig:fig3}.
Our plots indicate that the star formation front is self-perpetuating for $\qimf\fsn\gtrsim10^{-4}$ and $\estar\gtrsim10^{-4}$ in dwarf galaxies. 
This indicates that for a massive IMF, the required energy to sustain the propagation of the shock can be as low as $\sim 10^{-4}$ of the energy produced by SN, consistent with numerical simulation \citep{whalen2008}.
The front's characteristic speed is $\sim 300-700\,\kms$ during the epoch of reionization.
This indicates that the star formation front can initiate starbursts in dwarf galaxies and supply the needed ionizing photons in the early Universe \citep{finkelstein2015}.
\section{Summary \& discussion}
\label{sec:section4}
We explored the ignition of star formation in low-mass halos by a self-sustaining star formation front along cosmic filaments in the early Universe.
The gaseous fuel in these most abundant low-mass halos can not turn to stars due to their low-metallicity and low $\tvir$.
During the front's passage through each halo, the swept-up shell is capable of cooling rapidly and fragmenting into cold clumps that form stars on a timescale shorter than the front's dynamical timescale.
The propagation of the star formation front is maintained by energy injection from SN explosions, in analogy with the propagation of a detonation shock in gunpowder.
Assuming two-dimensional cylindrical symmetry, we find that the front traverses a filament with an average speed of $\sim 300-700\,\kms$.

As the star formation front propagates, the active region would appear to have a length of $\sim D_0\tstar\sim1.5$ kpc (corresponding to $\sim0.3''$ at $z\sim10$, resolvable by $JWST$), where $\tstar\sim3\times10^6$ yrs is the lifetime of massive stars ($\gtrsim10\,\msun$) which should dominate the UV emission in the early Universe \citep{bromm2011, loeb2013}.
We find that the length of the illuminated starburst region is up to ten times longer than the width of the filament at $z\gtrsim20$ for H$_2$ halos.
Future observations with the $JWST$ may reveal these elongated structures and constrain the speed of the star formation fronts.

Detection of the rest-frame UV flux from the illuminated fragment of filaments will constrain the free parameters of the detonation model, since the UV flux is correlated with the SFR, SFR $\approx1.4\, L_{\nu,28}\,\msun\,\rm yr^{-1}$, where $L_{\nu,28}=(L_{\nu}/10^{28}\,\rm erg\,s^{-1}\, Hz^{-1})$ is the UV luminosity at a rest-frame wavelength of $\sim 1250-1500$ \AA \citep[see][p.352]{loeb2013}.
Additionally, radio emission from the relativistic electrons produced in SN remnants can be measured to infer the SFR, as the SN rate tracks to the production rate of massive stars.
Our model assumes that SFR is proportional to the locally observed KS law with a correction factor $\estar$.
We find that the detonation mode of star formation fronts is viable for a SFR up to $\gtrsim 10^4$ times less efficient than associated with the KS law, indicating that even at the beginning of reionization, low-mass halos may experience starburst activity during the passage of a shock from triggered star formation in neighboring halos.
Radiative pre-processing by H$\,\textsc{ii}$ regions may be an additional source of energy injection to sustain the star formation fronts.
However, H$\,\textsc{ii}$ regions produce shocks of speed $\sim 30\,\kms$ in primordial halos \citep{wise2012}, much smaller than that produced by SN, which are the dominate energy source.
This self-sustaining mode of star formation fronts may account for the ionizing photons in low-mass halos at $z\gtrsim10$, as required by current observations \citep{robertson2015, anderson2017}.
Future probes of the faint end slope of the UV luminosity function of the star-forming galaxies with $JWST$ will be able to test our predictions for star forming fronts in cosmic filaments.
Even if the luminosity of an individual low-mass galaxy is below the detection threshold of $JWST$, filaments could be detectable since they contain many such galaxies.
%
%
\acknowledgements
We thank Simon D. Watt and Anna Rosen for useful discussions.
This work was supported in part by the Black Hole Initiative at Harvard University through a grant from the John Templeton Foundation.

\appendix
\section{Detonation model}
\label{sec:appendix}
In a one-dimensional laminar flow detonation model, there is a unique solution of $D_0$ corresponding to the minimum detonation speed that satisfies the conservation laws, known as the Chapman-Jouguet (CJ) velocity, $\dcj$ \citep{fickett1979, watt2012}.
For a two-dimensional cylindrical geometry (see Fig. \ref{fig:fig2}), we introduce a compressible streamline function, $\psi$, such that:
\begin{equation}
\left(\frac{\partial \psi}{\partial r}\right)_z=-r\rho v\;,\quad
\left(\frac{\partial \psi}{\partial z}\right)_r=r\rho u\;,
\end{equation}
Curves of constant $\psi$ are streamlines.
We can transform $(r,z)$ to a streamline based coordinate $(\psi,z)$, where $r=r(\psi,z)$. 
Eq. (\ref{eq:subeqns}) becomes:
\begin{subequations}
\label{eq:meqn}
\begin{gather}
\frac{\partial v}{\partial z}\left[v^2\left(1+\left(\frac{\partial r}{\partial z}\right)^2\right)-\cs^2\right]=\cs^2 v\left[\frac{\partial r}{\partial z}\frac{1}{r}+\frac{\partial^2 r}{\partial z \partial \psi}\left(\frac{\partial r}{\partial \psi}\right)^{-1}\right]-v^3\frac{\partial r}{\partial z}\frac{\partial^2 r}{\partial z^2}-(\gamma-1)QW\;,
\\
\frac{\partial\lambda}{\partial z}=\frac{W}{v}\;.
\end{gather}
\end{subequations}

If the shape of the streamlines, $r(\psi,z)$, were known a priori, then Eq. (\ref{eq:meqn}) reduce to a pair of ordinary differential equations for $v$ and $\lambda$ along each streamline where $\psi=$constant.
We assume that streamlines are straight but diverging, with their shape expressed as $r=\rf+F(\psi)(z-\zf)$.
Here $(\rf,\zf)$ denotes the shock front locus, and
\begin{equation}
F(\psi)=\left(\frac{\partial r}{\partial z}\right)_{\rm f}=\frac{\uf}{\vf}\;,
\end{equation}
where $\uf$ and $\vf$ are the post-shock flow velocities subject to shock jump conditions:
\begin{subequations}
\label{eq:shockjump}
\begin{equation}
\uf=-\frac{2D_0\zp}{\gamma+1}\left[1+(\zp)^2\right]^{-1}\;,
\end{equation}
\begin{equation}
\vf=-D_0\left[(\zp)^2+\frac{\gamma-1}{\gamma+1}\right]\left[1+(\zp)^2\right]^{-1}\;,
\end{equation}
\end{subequations}
where $\zp=d\zf/d\rf$.
Streamlines ahead of the shock are parallel, and thus $\psi=\rf^2\rho_0 D_0/2$, where $\rho_0$ is the density of the ambient medium.
Therefore, the solutions of Eq. (\ref{eq:meqn}) depend on the shock locus and shape via $(\rf,\zf,\zp,\zpp)$.
This results in an eigenvalue problem of $\zpp$, in that if $\zp$ was known a priori, there would be a unique $\zpp$ that satisfies Eq. (\ref{eq:shockjump}) and the CJ conditions \citep{fickett1979} for a given $D_0$:
\begin{equation}
v^2\left[1+\left(\frac{\partial x}{\partial y}\right)^2\right]-\cs^2=0\;,
\quad 
\left(\frac{\partial x}{\partial \psi}\right)^{-1}\cs^2 v\frac{\partial^2 x}{\partial y\partial \psi}-(\gamma-1)QW=0\;.
\end{equation}
We find that $D_0\propto1/\rvir$ due to two-dimensional effect, known as the diameter effect of detonation waves \citep{watt2012}.
%

%

\begin{thebibliography}{99}
%
\bibitem[Anderson et al.(2017)]{anderson2017} Anderson, L., Governato, F., Karcher, M., Quinn, T., \& Wadsley, J.\ 2017, \mnras, 468, 4077 
%
\bibitem[Arata et al.(2018)]{arata2018} Arata, S., Yajima, H., \& Nagamine, K.\ 2018, \mnras, 475, 4252 
%
\bibitem[Behroozi et al.(2013)]{behroozi2013} Behroozi, P.~S., Wechsler, R.~H., \& Conroy, C.\ 2013, \apj, 770, 57 
%
\bibitem[Behroozi \& Silk(2015)]{behroozi2015} Behroozi, P.~S., \& Silk, J.\ 2015, \apj, 799, 32 
%
\bibitem[Bond et al.(1996)]{bond1996} Bond, J.~R., Kofman, L., \& Pogosyan, D.\ 1996, \nat, 380, 603 
%
\bibitem[Bouwens et al.(2012)]{bouwens2012} Bouwens, R.~J., Illingworth, G.~D., Oesch, P.~A., et al.\ 2012, \apj, 754, 83 
%
\bibitem[Bouwens et al.(2015)]{bouwens2015} Bouwens, R.~J., Illingworth, G.~D., Oesch, P.~A., et al.\ 2015, \apj, 811, 140
%
\bibitem[Bromm \& Yoshida(2011)]{bromm2011} Bromm, V., \& Yoshida, N.\ 2011, \araa, 49, 373 
%
\bibitem[Chen et al.(2014)]{chen2014} Chen, P., Wise, J.~H., Norman, M.~L., Xu, H., \& O'Shea, B.~W.\ 2014, \apj, 795, 144
%
\bibitem[Dekel \& Silk(1986)]{dekel1986} Dekel, A., \& Silk, J.\ 1986, \apj, 303, 39 
%
\bibitem[Ferrara \& Scannapieco(2016)]{ferrara2016} Ferrara, A., \& Scannapieco, E.\ 2016, \apj, 833, 46
%
\bibitem[Fickett \& Davis(1979)]{fickett1979} Fickett, W., \& Davis, C.\ 1979, Los Alamos Series in Basic and Applied Sciences, Berkeley: University of California Press, 1979 
%
\bibitem[Field(1965)]{field1965} Field, G.~B.\ 1965, \apj, 142, 531 
%
\bibitem[Finkelstein et al.(2015)]{finkelstein2015} Finkelstein, S.~L., Ryan, R.~E., Jr., Papovich, C., et al.\ 2015, \apj, 810, 71
%
\bibitem[Furlanetto \& Loeb(2003)]{furlanetto2003} Furlanetto, S.~R., \& Loeb, A.\ 2003, \apj, 588, 18
%
\bibitem[Haiman et al.(1996)]{haiman1996} Haiman, Z., Thoul, A.~A., \& Loeb, A.\ 1996, \apj, 464, 523
%
\bibitem[Hopkins et al.(2011)]{hopkins2011} Hopkins, P.~F., Quataert, E., \& Murray, N.\ 2011, \mnras, 417, 950 
%
\bibitem[Inoue \& Omukai(2015)]{inoue2015} Inoue, T., \& Omukai, K.\ 2015, \apj, 805, 73
%
\bibitem[Kennicutt(1998)]{kennicutt1998} Kennicutt, R.~C., Jr.\ 1998, \apj, 498, 541 
%
\bibitem[King \& Pounds(2015)]{king2015} King, A., \& Pounds, K.\ 2015, \araa, 53, 115 
%
\bibitem[Krumholz \& Dekel(2012)]{krumholz2012} Krumholz, M.~R., \& Dekel, A.\ 2012, \apj, 753, 16 
%
\bibitem[Maio et al.(2007)]{maio2007} Maio, U., Dolag, K., Ciardi, B., \& Tornatore, L.\ 2007, \mnras, 379, 963 
%
\bibitem[Maiolino et al.(2017)]{maiolino2017} Maiolino, R., Russell, H.~R., Fabian, A.~C., et al.\ 2017, \nat, 544, 202
%
\bibitem[Martin et al.(2012)]{martin2012} Martin, C.~L., Shapley, A.~E., Coil, A.~L., et al.\ 2012, \apj, 760, 127
%
\bibitem[Mo et al.(2010)]{mo2010} Mo, H., van den Bosch, F.~C., \& White, S.\ 2010, Galaxy Formation and Evolution, Cambridge, UK: Cambridge University Press, 2010
%
\bibitem[Mori et al.(2002)]{mori2002} Mori, M., Ferrara, A., \& Madau, P.\ 2002, \apj, 571, 40 
%
\bibitem[Loeb \& Furlanetto (2013)]{loeb2013} Loeb, A. \& Furlanetto, S. R., \ 2013, The First Galaxies in the Universe, Princeton University Press.
%
\bibitem[Peeples \& Shankar(2011)]{peeples2011} Peeples, M.~S., \& Shankar, F.\ 2011, \mnras, 417, 2962
%
\bibitem[Press \& Schechter(1974)]{press1974} Press, W.~H., \& Schechter, P.\ 1974, \apj, 187, 425
%
\bibitem[Robertson et al.(2015)]{robertson2015} Robertson, B.~E., Ellis, R.~S., Furlanetto, S.~R., \& Dunlop, J.~S.\ 2015, \apjl, 802, L19 
%
\bibitem[Rupke et al.(2005)]{rupke2005} Rupke, D.~S., Veilleux, S., \& Sanders, D.~B.\ 2005, \apj, 632, 751
%
\bibitem[Scannapieco(2017)]{scannapieco2017} Scannapieco, E.\ 2017, \apj, 837, 28
%
\bibitem[Schaye \& Dalla Vecchia(2008)]{schaye2008} Schaye, J., \& Dalla Vecchia, C.\ 2008, \mnras, 383, 1210 
%
\bibitem[Schneider et al.(2018)]{schneider2018} Schneider, E.~E., Robertson, B.~E., \& Thompson, T.~A.\ 2018, arXiv:1803.01005
%
\bibitem[Sheth \& Tormen(1999)]{sheth1999} Sheth, R.~K., \& Tormen, G.\ 1999, \mnras, 308, 119
%
\bibitem[Silk(1997)]{silk1997} Silk, J.\ 1997, \apj, 481, 703 
%
\bibitem[Silk(2013)]{silk2013} Silk, J.\ 2013, \apj, 772, 112 
%
\bibitem[Strickland \& Heckman(2009)]{strickland2009} Strickland, D.~K., \& Heckman, T.~M.\ 2009, \apj, 697, 2030 
%
\bibitem[Sturm et al.(2011)]{sturm2011} Sturm, E., Gonz{\'a}lez-Alfonso, E., Veilleux, S., et al.\ 2011, \apjl, 733, L16
%
\bibitem[Sutherland \& Dopita(1993)]{sutherland1993} Sutherland, R.~S., \& Dopita, M.~A.\ 1993, \apjs, 88, 253 
%
\bibitem[Thompson et al.(2016)]{thompson2016} Thompson, T.~A., Quataert, E., Zhang, D., \& Weinberg, D.~H.\ 2016, \mnras, 455, 1830
%
\bibitem[Tremonti et al.(2007)]{tremonti2007} Tremonti, C.~A., Moustakas, J., \& Diamond-Stanic, A.~M.\ 2007, \apjl, 663, L77
%
\bibitem[Trenti \& Stiavelli(2009)]{trenti2009} Trenti, M., \& Stiavelli, M.\ 2009, \apj, 694, 879
%
\bibitem[Wang \& Loeb(2015)]{wang2015} Wang, X., \& Loeb, A.\ 2015, \mnras, 453, 837
%
\bibitem[Wang \& Loeb(2018)]{wang2018} Wang, X., \& Loeb, A.\ 2018, New Astro., 61, 95 
%
\bibitem[Watt et al.(2012)]{watt2012} Watt, S.~D., Sharpe, G.~J., Falle, S.~A.~E.~G., \& Braithwaite, M.\ 2012, Journal of Engineering Mathematics, 75, 1
%
\bibitem[Whalen et al.(2008)]{whalen2008} Whalen, D., van Veelen, B., O'Shea, B.~W., \& Norman, M.~L.\ 2008, \apj, 682, 49 
%
\bibitem[Wise et al.(2012)]{wise2012} Wise, J.~H., Abel, T., Turk, M.~J., Norman, M.~L., \& Smith, B.~D.\ 2012, \mnras, 427, 311
%
\bibitem[Wise et al.(2014)]{wise2014} Wise, J.~H., Demchenko, V.~G., Halicek, M.~T., et al.\ 2014, \mnras, 442, 2560 
%
\bibitem[Zubovas \& King(2014)]{zubovas2014} Zubovas, K., \& King, A.~R.\ 2014, \mnras, 439, 400
%
\end{thebibliography}
\end{document}